\pdfoutput=1

\documentclass[11pt]{article}

\usepackage[preprint]{acl}

\usepackage{times}
\usepackage{latexsym}
\usepackage{booktabs}
\usepackage{amsmath}
\usepackage{amssymb}
\usepackage{multirow}
\usepackage[T1]{fontenc}

\usepackage[utf8]{inputenc}

\usepackage{microtype}

\usepackage{inconsolata}

\usepackage{graphicx}
\usepackage{xcolor}
\usepackage{algorithm}
\usepackage{algpseudocode}
\usepackage{tikz}
\usetikzlibrary{positioning, arrows.meta, shapes.geometric, fit, backgrounds}
\usepackage{pgfplots}
\pgfplotsset{compat=1.18}
\usepackage{enumitem}

\newcommand{\xhdr}[1]{{\noindent\bfseries #1}.} 

\title{EventConnector: Mining Social Event Relations through Temporal Graphs}

\author{
\begin{tabular}{c}
\textbf{Zijie Lei\textsuperscript{1}} \quad
\textbf{Haofei Yu\textsuperscript{2}} \quad
\textbf{Ge Liu\textsuperscript{2}} \quad
\textbf{Jiaxuan You\textsuperscript{2}} \\
\\[-0.6em]
\textsuperscript{1}Meta Monetization AI \\
\textsuperscript{2}University of Illinois Urbana-Champaign
\end{tabular}
}

\begin{document}
\maketitle

\begin{abstract}
Understanding and retrieving related real-world events based on their temporal dynamics is a fundamental challenge in time-sensitive applications such as forecasting, information retrieval, and social analysis. Existing methods often rely on semantic similarity or global time-series alignment, which overlook the transient and directional dependencies that frequently underlie real-world correlations. In this work, we introduce \textit{EventConnector}, a framework that constructs a temporal event graph capturing localized co-fluctuations and lead-lag relationships between events through their time-series trajectories. We further propose \textbf{EC-Fusion}, an adaptive retrieval mechanism that fuses EventConnector's graph-based scores with a complementary Granger-causal signal via a graph-quality-aware mixing weight. Across two real-world prediction market benchmarks (Polymarket and Kalshi) and nine forecasting architectures evaluated over three random seeds, EC-Fusion is the best non-oracle retrieval method on $17/18$ model--dataset cells, reducing RMSE by $6.87\%$ on average (up to $10.86\%$) over the strongest comparable retrieval baseline, with statistical significance at $p < 0.01$ after Holm--Bonferroni correction. These results highlight the effectiveness of temporally grounded graph modeling, augmented with causal-signal fusion, in capturing latent event relationships beyond what semantic similarity or traditional alignment techniques can offer.
\end{abstract}


\section{Introduction}
Real-world events rarely unfold in isolation—they are embedded within interdependent systems spanning political, economic, and cultural domains~\citep{von1968general, kelley1959interdependence}. Modeling the temporal dependencies among such events is crucial not only for forecasting, but also for understanding how societal processes co-evolve through mechanisms like information diffusion~\citep{rogers2003diffusion, myers2012information}. For example, a major fiscal policy announcement can ripple through financial markets, as evidenced by studies linking President Trump’s public statements to fluctuations in cryptocurrency prices~\citep{Huynh2021Bitcoin}. Anticipating these cross-domain ripple effects is essential for informing public policy, risk assessment, and strategic decision-making~\citep{benzinevent2023}, yet it remains a challenging and underexplored problem.

\begin{figure}[t]
\centering
\includegraphics[width=\linewidth]{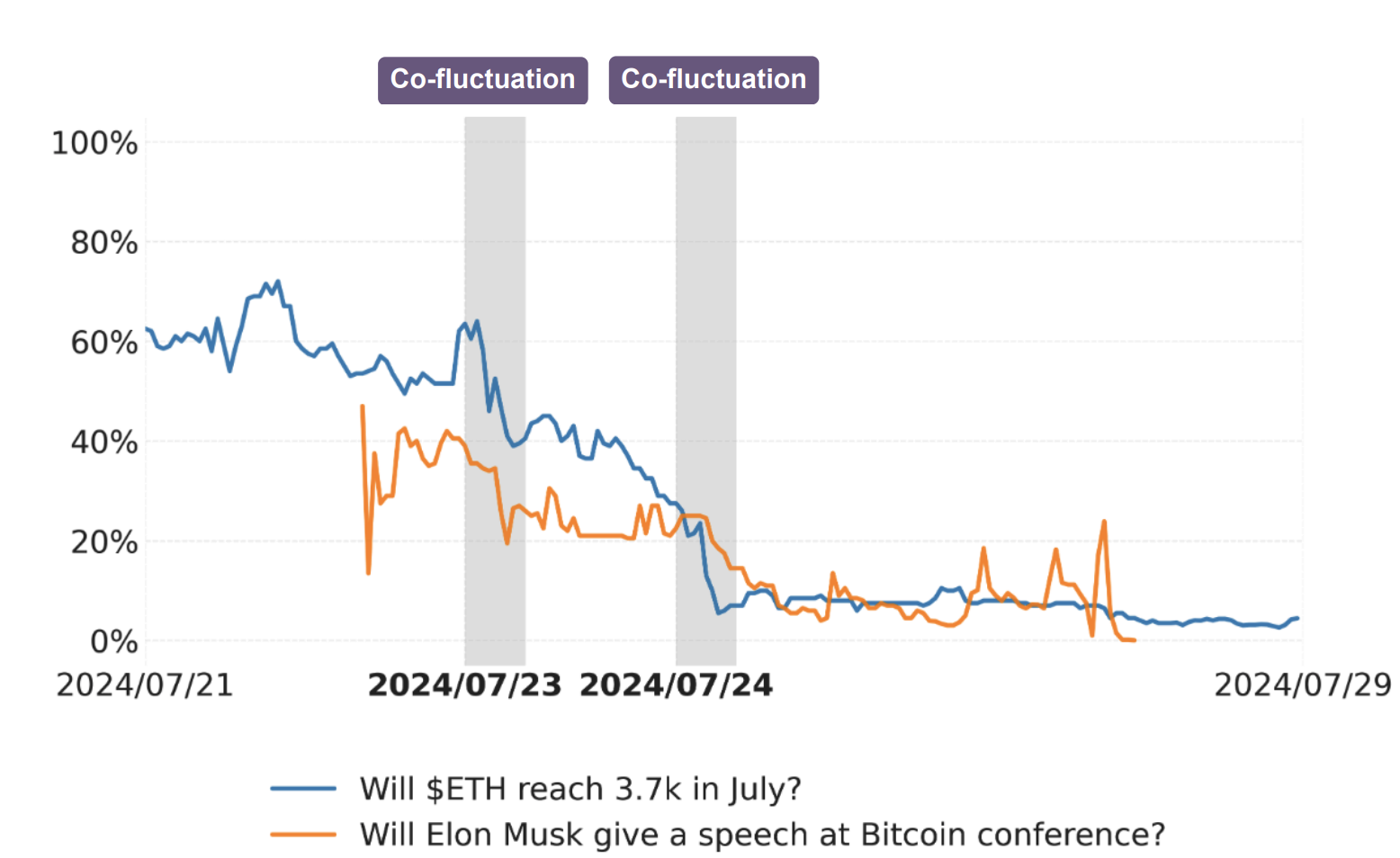}
\caption{\textbf{Semantically unrelated events can exhibit co-fluctuations and be secretly connected.} The two Polymarket events---\textit{``Will \$ETH reach 3.7k in July?''} (blue) and \textit{``Will Elon Musk give a speech at the Bitcoin conference?''} (orange)---are semantically unrelated, yet their market-implied probabilities track each other closely. Such short-term alignment reflects shared sentiment or speculative drivers that traditional semantic similarity cannot capture.}
\label{fig:non-semantic-correlation}
\end{figure}

\vspace{1mm}
\xhdr{Challenges for finding event connection} One major obstacle is that correlated events are often \emph{semantically dissimilar}: a shift in trade policy may temporally align with cryptocurrency-price movements without any lexical or ontological similarity to suggest a connection (Figure~\ref{fig:non-semantic-correlation}). Traditional tools---Granger causality~\citep{Granger1969} or Hawkes processes~\citep{Hawkes1971}---assume linearity or low-noise environments and struggle on non-stationary event-driven signals~\citep{shojaie2022granger, jaisson2015limit}. Modern alternatives are no better fit: deep forecasting models assume a shared latent space or predefined relational structure (well-suited to homogeneous systems like traffic networks~\citep{Li2018DCRNN}, less so for heterogeneous events~\citep{lim2021time}), while temporal knowledge graph methods~\citep{Goel2020Diachronic, Han2020Subgraph, cai2023temporal} embed all entities into a unified space, risk collapsing structurally distinct signals, and typically assume stable dynamics~\citep{kazemi2020representation, trivedi2019dyrep}.

\vspace{1mm}
\xhdr{Connecting events with a social temporal graph and causal fusion} In this work, we present \textit{EventConnector}, a temporal graph framework for discovering and modeling dynamic dependencies between social events based on their evolving time series. Our method constructs a data-driven \textit{social temporal graph} in which nodes represent individual events and edges encode localized, statistically significant relationships derived from short-term co-fluctuation and lead-lag inference. To exploit complementary causal structure that pure co-fluctuation may miss, we additionally introduce \textbf{EC-Fusion}: a parallel directed graph $\mathcal{G}_{\text{GC}}$ built from pairwise Granger-causality tests is combined with the EC graph through an adaptive, graph-quality-aware mixing weight. Together, the two graphs form a structurally and causally aware retriever that surfaces non-obvious cross-domain associations, supports multi-hop reasoning over indirect chains of dependency, and provides inductive bias to any downstream forecasting architecture without modifying its training procedure.

\vspace{1mm}
\xhdr{Key contributions} We make four contributions. (i)~We formalize the \emph{social temporal graph} and an inductive retrieval pipeline (anchor selection $\to$ multi-hop BFS expansion $\to$ training-window augmentation) that decouples retrieval from the forecasting model. (ii)~We propose \textbf{EC-Fusion}, an adaptive fusion of the co-fluctuation graph with a Granger-causal graph, parameterized by a graph-quality-aware mixing weight $\alpha$ that requires no per-query tuning at inference. (iii)~We introduce a \emph{hybrid anchor} that combines price- and text-based similarity and quantitatively dominates either modality alone. (iv)~We evaluate on \emph{two} datasets (Polymarket and Kalshi) and \emph{nine} forecasting architectures across three random seeds, demonstrating that EC-Fusion is the best non-oracle method on $17/18$ model--dataset pairs and that its retrievals are almost entirely disjoint from those of text-similarity baselines (mean Jaccard $0.016$; $70\%$ of queries have zero question overlap), evidencing complementary signal that text retrieval cannot recover.


\section{Related Works}

\xhdr{Time-series event modeling} Classical techniques like Dynamic Time Warping (DTW) \cite{Berndt1994}, local correlation tracking \cite{Papadimitriou2006}, and BRAID \cite{Sakurai2005} align or group time series based on transient or lagged patterns. Matrix profile methods \cite{Yeh2016} efficiently detect similar or anomalous subsequences. Directional dependencies are modeled through high-dimensional Granger causality \cite{Arnold2007} and lead-lag networks \cite{Bennett2022}. Point-process models like multidimensional Hawkes processes capture self-/cross-exciting dynamics \cite{Zhou2013}. Deep forecasting models (e.g., LSTNet \cite{Lai2018}) leverage convolutional and recurrent layers for multiscale temporal dependencies. These approaches inform EventConnector's use of co-fluctuation and causality for graph construction.
\vspace{1mm}

\xhdr{Society system modeling and social event forecasting} Foundational models—DeGroot averaging \cite{DeGroot1974}, threshold-based diffusion \cite{Granovetter1978}, and bounded-confidence dynamics \cite{Hegselmann2002}—explain macro patterns from individual behavior. Data-driven systems such as EMBERS \cite{Ramakrishnan2014}, spatio-temporal forecasting \cite{Zhao2015}, and nested MIL \cite{Ning2016} infer emergent trends from open signals. Evolving event-context graphs \cite{deng2019graph} further capture event interplay. Earlier social event forecasting methods used social media and statistical signals (e.g., scan statistics \cite{chen2014nonparametric}, cascade models \cite{cadena2015forecasting}) to forecast unrest. Temporal event chains \cite{radinsky2012modeling} and entity-centric graph models \cite{deng2019graph, deng2020glean, li2021temporal} incorporate cause-effect and multimodal dynamics. EventConnector builds on these ideas, defining a temporal event graph grounded in co-fluctuation and lead-lag signals for structure-aware forecasting.
\vspace{1mm}

\xhdr{Temporal graphs} Temporal GNNs like TGAT \cite{xu2020inductive}, DySAT \cite{sankar2020dysat}, Know-Evolve \cite{trivedi2017know}, and DyRep \cite{trivedi2019dyrep} embed evolving node relations via time-aware attention or event-driven dynamics. Message-passing models like TeMP \cite{wu2020temp} propagate over time-stamped knowledge graphs. EventConnector differs by defining temporal edges from time-series co-fluctuation and directional influence, enabling both inductive retrieval and forecasting.
\vspace{1mm}

\xhdr{Retrieval augmentation and score fusion} Retrieval augmentation grounds parametric models in non-parametric memory, with origins in open-domain QA (DPR, ColBERT~\citep{karpukhin2020dpr,khattab2020colbert}) and retrieval-augmented generation~\citep{lewis2020retrieval}, and recent extensions to time-series forecasting~\citep{woo2024unified} and temporal knowledge-graph QA~\citep{qian2024timer4}. A parallel line of work on hybrid sparse--dense retrieval has shown that combining a dense encoder (e.g., SBERT~\citep{reimers2019sentence}) with a classical lexical retriever (BM25~\citep{robertson2009bm25}) via Reciprocal Rank Fusion~\citep{cormack2009rrf} outperforms either signal alone. EC-Fusion is an analogue of this design pattern in the temporal domain: the EC graph captures local co-fluctuation (analogous to dense match) and the Granger graph captures asymmetric predictive structure (analogous to a complementary causal signal). Unlike RRF, our mixing weight $\alpha$ is computed from structural diagnostics of the retriever, so the fusion adapts to dataset characteristics without requiring a held-out tuning set.


\section{Background}
\label{sec:background}

\xhdr{Prediction markets as a data source} Prediction markets aggregate dispersed information into collective forecasts about uncertain future outcomes~\citep{wolfers2004}, with market-implied prices interpretable as consensus probabilities that have outperformed traditional polling in U.S.\ election forecasting~\citep{rothschild2009}. We use two complementary platforms: \textit{Polymarket}, a cryptocurrency-funded market dominated by politics, elections, and crypto questions, and \textit{Kalshi}, a CFTC-regulated U.S.\ exchange with a more diverse mix of economic, policy, and entertainment outcomes. The two differ in scale and market structure, making a method that succeeds on both more credibly capturing event dynamics than platform-specific behavior.

\xhdr{Social events} We denote by $\mathcal{E} = \{e_1, e_2, \dots, e_N\}$ a collection of real-world \textit{social events}, where each event $e_i$ corresponds to a temporally evolving question or proposition about the world (e.g., ``Will a political candidate win the election?'' or ``Will Bitcoin reach \$40{,}000 by next month?''). Formally, we define a social event as a tuple
\begin{equation}
\label{event-equation}
    e = (q, \mathcal{O}),
\end{equation}
where $q$ is a future-uncertain question and $\mathcal{O} = \{o_1, \dots, o_K\}$ is a set of $K$ mutually exclusive outcomes. Each outcome $o_k \in \mathcal{O}$ is associated with a probability time series $\{p_k(t)\}_{t=1}^T$, with $p_k(t) \in [0, 1]$ denoting the market-implied probability of $o_k$ at time step $t$. Outcome probabilities are normalized at each timestamp: $\sum_{k=1}^K p_k(t) = 1$ for all $t \in \{1,\dots,T\}$. The trajectory $\{p_k(t)\}$ thus serves as an indirect, behaviorally grounded observation of the event itself: it encodes how public belief about $o_k$ shifts over time in response to campaign events, polling results, media coverage, price action, or macroeconomic signals.

\xhdr{Connections between social events} Two events $e_i$ and $e_j$ are \emph{connected} when their belief trajectories exhibit statistically meaningful co-evolution---synchronized fluctuations, lead--lag alignment, or recurring directional response---irrespective of topical or lexical similarity between $q_i$ and $q_j$. Discovering and exploiting these belief-dynamics-based connections is the task this paper addresses.


\section{Problem Formulation}
\label{sec:formulation}

\xhdr{Forecasting objective}
Given the historical belief trajectories $\{p_k(1),\dots,p_k(T)\}$ for each outcome $k\in\{1,\dots,K\}$ of a query event $e_q$, the forecasting objective is to predict the next $H$ values $\{p_k(T+1),\dots,p_k(T+H)\}$. This is a multi-horizon time-series prediction problem: anticipate the evolution of collective belief under ongoing information flow. The formulation is platform-agnostic: any prediction market that records belief-evolving probabilities for mutually exclusive outcomes admits the same definition. We instantiate it on two platforms (Polymarket and Kalshi) with disjoint event sets, building one social temporal graph per dataset; no cross-platform information is shared between graphs.

\xhdr{Temporal graph}
A \textit{temporal graph} is a tuple $\mathcal{G}_T = (\mathcal{V}, \mathcal{E}_T)$ where $\mathcal{V}$ is a set of nodes and $\mathcal{E}_T \subseteq \mathcal{V} \times \mathcal{V} \times \mathbb{R}_+$ is a set of time-stamped edges. Each edge $(u, v, t) \in \mathcal{E}_T$ indicates an interaction between nodes $u$ and $v$ active at time $t$ (or over an interval). Unlike static graphs, $\mathcal{G}_T$ encodes \textit{when} connections occur, supporting analysis of causality, influence propagation, and dynamic neighborhood evolution.

\xhdr{Social temporal graph}
We specialize $\mathcal{G}_T$ to the social domain: each node $v_i \in \mathcal{V}$ represents a unique social event $e_i = (q_i, \mathcal{O}_i)$ with its belief-trajectory time series, and each edge $(v_i, v_j, t) \in \mathcal{E}_T$ encodes a time-specific correlation or influence between two events, inferred from their respective trajectories. This is the formal object on which the EventConnector retriever (Sec.~\ref{sec:methods}) operates.

\xhdr{Key properties of the social temporal graph}
Three properties distinguish $\mathcal{G}_T$ from semantic or static-graph alternatives. \emph{(i)~Non-semantic edges:} edges are induced exclusively from observed co-fluctuation in belief trajectories---synchronous surges, lead--lag alignment, or recurring directional response---rather than from textual similarity between the questions $q_i, q_j$. \emph{(ii)~Cross-domain reach:} because the edge criterion is behavioral, $\mathcal{G}_T$ surfaces latent dependencies that span semantically distant topics (e.g., linking a fiscal-policy event with a cryptocurrency-price event when both respond to a common macro shock). \emph{(iii)~Inductive grounding for downstream tasks:} $\mathcal{G}_T$ provides a relational substrate that supports inductive query handling for unseen events, structured retrieval of training context, and forecast augmentation under sparse supervision. Empirical statistics of the instantiated graph on each dataset are summarized in Table~\ref{tab:graph-stats}.

\begin{table}[t]
\centering
\small
\caption{\textbf{Social Temporal Graph statistics} (built from a $N_{\mathrm{STG}}{=}500$ subsample, $\tau_{\mathrm{corr}}{=}0.7$, max 5 edges/node).}
\label{tab:graph-stats}
\setlength{\tabcolsep}{4pt}
\begin{tabular}{lcc}
\toprule
\textbf{Metric} & \textbf{Polymarket} & \textbf{Kalshi} \\
\midrule
Canonical nodes (post-merge) & 155 & 190 \\
Edges & 612 & 684 \\
Density & 0.051 & 0.038 \\
Avg.\ degree & 7.90 & 7.20 \\
Max degree & 24 & 18 \\
Connected components & 1 & 1 \\
Co-fluctuation edges & 597 (97.5\%) & 684 (100.0\%) \\
DTW enrichment edges & 15 (2.5\%) & 0 (0.0\%) \\
Mean edge weight & 0.971 & 0.968 \\
Synchronous edges (lag $=0$) & 28.1\% & 27.0\% \\
\bottomrule
\end{tabular}
\end{table}


\section{EventConnector: Connecting Events with Social Temporal Graphs}
\label{sec:methods}

Our framework operates in three stages: (1)~\textbf{Construct} the social temporal graph from training events, (2)~\textbf{Retrieve} relevant events for a new query via inductive graph search, and (3)~\textbf{Predict} by training a forecasting model on the retrieved neighborhood. The complete procedure is given in Algorithm~\ref{alg:ecfusion}; we describe each stage below.

\subsection{Stage 1: Graph Construction}
\label{sec:stage1}

Given a pool of $N$ training events, each with a price time series $\{p(t)\}_{t=1}^T$, we construct the Social Temporal Graph $\mathcal{G} = (\mathcal{V}, \mathcal{E})$ through three sequential steps.

\xhdr{Step 1: Node Construction via Event Merging}
To prevent near-duplicate events (e.g., rephrasings of the same question) from inflating neighborhoods and reducing retrieval diversity, we consolidate them via union-find: for every pair whose full price series exhibit Pearson correlation $|r| > \tau_{\text{merge}}$ ($\tau_{\text{merge}} = 0.95$), we merge them into a single canonical node that inherits the representative member's price series. This reduces $N$ raw events to $|\mathcal{V}|$ canonical nodes.

\xhdr{Step 2: Co-Fluctuation Edge Construction}
For each pair of canonical nodes $(v_i, v_j)$, we compute sliding-window Pearson correlation over their price series using a window of size $w$ (default $w = 7$ for daily data). Let $\rho_{ij}^{(s)}$ denote the correlation in the window starting at position $s$. If $\max_s |\rho_{ij}^{(s)}| > \tau_{\text{corr}}$ (we use $\tau_{\text{corr}} = 0.7$), we create a temporal edge $(v_i, v_j)$ with weight equal to the maximum observed correlation. We additionally store the lag $\ell^* = \arg\max_{\ell \in [-L, L]} |r_{xy}(\ell)|$ ($L=3$) as edge metadata for downstream lead-lag analyses.

\xhdr{Step 3: Multi-Hop Enrichment via DTW}
Direct co-fluctuation edges may miss events that exhibit similar temporal shapes but with different timing or amplitude. For pairs of canonical nodes not connected by a co-fluctuation edge, we compute the Dynamic Time Warping (DTW) distance between their min-max normalized price series, using a Sakoe-Chiba band constraint for efficiency. If $d_{\text{DTW}}(v_i, v_j) < \tau_{\text{DTW}}$ (default $\tau_{\text{DTW}} = 1.5$), we add a second-order enrichment edge with weight $1 - d_{\text{DTW}} / \tau_{\text{DTW}}$. This step captures transitive temporal dependencies that complement the direct co-fluctuation structure.

\xhdr{Subsampling for Tractability}
Edge construction is $\mathcal{O}(N^2)$, so we build the STG on a uniformly subsampled set of $N_{\text{STG}} = 500$ records per dataset; retrieval at inference time expands BFS over the resulting canonical-node graph, and retrieved indices are mapped back to all merged member records before forecasting.

\subsection{Stage 2: Inductive Retrieval}
\label{sec:stage2}

Given a new query event $e_q$ with input prices $\{p_q(t)\}_{t=1}^{T_{\text{in}}}$, the retrieval stage maps it onto the graph and collects relevant training events. Crucially, only the \emph{input} portion of the query's price series is used for anchor selection, preventing data leakage from the prediction target.

\xhdr{Anchor Selection}
The \emph{anchor node} $v^*$ is selected by maximizing a \textbf{hybrid score} that combines a price-correlation term with an SBERT~\cite{reimers2019sentence} textual similarity: $\text{score}(v) = (1 - \beta) \cdot |r(p_q, p_v)| + \beta \cdot \cos(\phi(q_q), \phi(q_v))$, where $\phi(\cdot)$ is the SBERT \texttt{all-MiniLM-L6-v2} encoder. We use $\beta = 0.3$ (70\% price, 30\% text) throughout, which empirically dominates either modality alone.

\xhdr{BFS Neighborhood Expansion}
Starting from the anchor $v^*$, we perform breadth-first search up to $n$ hops (default $n = 2$) on $\mathcal{G}$. Each visited node is scored by the \emph{minimum edge weight along the path from the anchor} (bottleneck path weight), ensuring that weakly connected nodes receive lower scores even if they are nearby in hop distance.

\xhdr{Candidate Ranking}
All training records mapped to the visited canonical nodes (including merged records) are ranked by their canonical node's BFS score. The top-$k$ records are returned as the retrieved training set.

\subsection{Stage 2b: EC-Fusion --- Adaptive Granger-Causal Fusion}
\label{sec:ecfusion}

The co-fluctuation graph captures synchronous and lagged correlations, but may miss purely \emph{causal} relationships where one event's past values predict another's future without exhibiting similar price shapes. To address this, we introduce \textbf{EC-Fusion}, which adaptively combines EventConnector's graph-based retrieval with a complementary Granger-causal retrieval signal. The full procedure is given in Algorithm~\ref{alg:ecfusion}.

\begin{algorithm}[t]
\caption{EC-Fusion retrieval}
\label{alg:ecfusion}
\small
\begin{algorithmic}[1]
\Require query $e_q$, training events $\mathcal{E}$, budget $k$
\Statex \emph{One-time, before all queries:}
\State Build STG $\mathcal{G}_T$ from $\mathcal{E}$ (merge, co-fluctuation edges, DTW enrichment)
\State Build Granger graph $\mathcal{G}_{\text{GC}}$ from $\mathcal{E}$ ($F$-test, $p<0.05$)
\State Compute graph quality $q$ and set $\alpha = \sigma(8(q-0.55))$ within $[\alpha_{\min}, \alpha_{\max}]$
\Statex \emph{Per query:}
\State Hybrid anchor $v^* \gets \arg\max_v (1{-}\beta)|r(p_q,p_v)| + \beta\cos(\phi(q_q),\phi(q_v))$
\State EC scores $s_{\text{EC}}$ via bottleneck BFS from $v^*$ on $\mathcal{G}_T$
\State Granger scores $s_{\text{GC}}$ via reciprocal rank on $\mathcal{G}_{\text{GC}}$
\State Fuse $s = \alpha\,\hat{s}_{\text{EC}} + (1{-}\alpha)\,\hat{s}_{\text{GC}}$ (min-max normalized)
\State \Return top-$k$ records by $s$
\end{algorithmic}
\end{algorithm}

\xhdr{Granger Causality Graph}
We construct a separate directed graph $\mathcal{G}_{\text{GC}}$ where an edge $(v_i \to v_j)$ exists if $v_i$ Granger-causes $v_j$ at significance level $p < 0.05$, tested via an F-test over lags $\{1, \ldots, L\}$ ($L = 3$). The graph is built over \emph{event-level aggregated} price series (concatenating all training windows per unique event) for sufficient statistical power, and we cap the Granger graph at the first $N_{\text{GC}} = 100$ event-aggregated series to keep the $\mathcal{O}(N^2)$ pairwise testing tractable; the EC graph $\mathcal{G}$ is unaffected and spans all canonical nodes.

\xhdr{Adaptive $\alpha$ Selection}
The mixing weight $\alpha$ is derived from a graph-quality score $q \in [0, 1]$ computed from four diagnostics of $\mathcal{G}$: \emph{dead-series ratio} (fraction of nodes with near-zero variance), \emph{density} (over-connected $\Rightarrow$ noisy), \emph{merge rate} (high $\Rightarrow$ redundant; the most heavily weighted signal), and \emph{average degree}; the exact weighted combination is given in Appendix~\ref{sec:appendix-hyperparams}. We then map $q$ to $\alpha$ via a sigmoid:
\begin{equation}
    \alpha = \alpha_{\min} + (\alpha_{\max} - \alpha_{\min}) \cdot \sigma(8 (q - 0.55))
\end{equation}
with $\alpha_{\min} = 0.2$, $\alpha_{\max} = 0.8$. Clean graphs receive higher $\alpha$ (trusting EC); noisier graphs receive lower $\alpha$ (leaning on Granger).

\subsection{Stage 3: Forecasting with Retrieved Context}
\label{sec:stage3}

For each of the top-$k$ retrieved events $e_i$, we extract non-overlapping sliding windows $(x, y)$ from its price series (with $x$ the input segment and $y$ the prediction target) and add them to the training set. The base forecasting model is then trained on these augmented samples with early stopping on a held-out validation split. Because retrieval is fully decoupled from the architecture, EventConnector plugs into any time-series forecaster.


\section{Experimental Settings}
\label{sec:experiments}

\xhdr{Datasets}
We evaluate on two real-world prediction market datasets from SWM-Bench~\citep{yu2026socialworldmodels}: \textbf{Polymarket}, a cryptocurrency-based platform spanning politics, elections, cryptocurrency, and other topics; and \textbf{Kalshi}, a CFTC-regulated U.S.\ exchange covering economics, politics, entertainment, and other topics. Both datasets provide event-level probability time series at daily resolution. We use a non-overlapping sliding window strategy with $T_{\text{in}} = 10$ input timesteps and $H = 7$ prediction horizon (i.e., 10 days of history $\to$ 7 days of forecast). Per-dataset statistics are summarized in Table~\ref{tab:dataset-stats}.

\begin{table}[t]
\centering
\small
\caption{\textbf{Dataset statistics.} Unique events count distinct market questions. Polymarket spans 2022-11 to 2026-01; Kalshi spans 2024-05 to 2025-12.}
\label{tab:dataset-stats}
\setlength{\tabcolsep}{5pt}
\begin{tabular}{lcc}
\toprule
\textbf{Statistic} & \textbf{Polymarket} & \textbf{Kalshi} \\
\midrule
Training windows & 9{,}803 & 2{,}779 \\
Test windows & 3{,}692 & 1{,}120 \\
Unique training events & 973 & 627 \\
Unique test events & 515 & 360 \\
Top categories & Politics, & Politics, \\
(\# train windows) & Election, & Entertainment, \\
& Crypto & Economics \\
\bottomrule
\end{tabular}
\end{table}

\xhdr{Forecasting Models}
We evaluate nine representative architectures spanning linear, MLP, Transformer, and CNN families:
\textit{DLinear}~\cite{zeng2023transformers},
\textit{N-BEATS}~\cite{oreshkin2019n},
\textit{TimesNet}~\cite{wu2022timesnet},
\textit{PatchTST}~\cite{nie2023time},
\textit{iTransformer}~\cite{liu2024itransformer},
\textit{Informer}~\cite{zhou2021informer},
\textit{Autoformer}~\cite{wu2021autoformer},
\textit{TSMixer}~\cite{chen2023tsmixer}, and
\textit{TiDE}~\cite{das2023tide}.
All models use publicly available implementations with hyperparameters tuned on a held-out validation set. Detailed model descriptions appear in Appendix~\ref{sec:appendix-b}.

\xhdr{Retrieval Baselines}
We compare EC-Fusion against five retrieval strategies:
(1)~\textit{Few-Shot}: $k$ randomly sampled training events (minimal context);
(2)~\textit{TimeSeries}: $k$-nearest neighbors by DTW distance over full price trajectories;
(3)~\textit{Semantic}: $k$-nearest by SBERT~\cite{reimers2019sentence} cosine similarity on event question text;
(4)~\textit{BM25}~\cite{robertson2009bm25}: $k$-nearest by BM25 lexical overlap on questions;
(5)~\textit{Category}: $k$ events from the same domain/category.
We additionally include \textit{EventConnector} (price-only anchor, no Granger fusion) and \textit{EC-Hybrid} (hybrid anchor and output scoring, no Granger) as design-ablation baselines; \textit{Full-Shot} trains on all available in-domain data and serves as an oracle upper bound.

\xhdr{Evaluation Protocol}
All experiments are repeated over 3 random seeds (42, 123, 7). We report mean$_{\pm\text{std}}$ RMSE and MAE. For EC-Fusion, we select the best fusion weight $\alpha$ per seed from $\{0.0, 0.3, 0.5, 0.7, 1.0, \text{adaptive}\}$ based on validation RMSE, then report the test-set performance at that $\alpha$. The main results, including the new mean$_{\pm\text{std}}$ Table~\ref{tab:main-results}, are presented in Section~\ref{sec:exp_results}.


\begin{table*}[!t]
\setlength\tabcolsep{4pt}
\centering
\small
\caption{\textbf{Forecasting performance (RMSE$_{\pm\text{std}}$, lower is better).}
Mean over 3 seeds (42, 123, 7). \textbf{Bold} = best non-oracle retrieval method by mean RMSE; \underline{underline} = second best. EC-Fusion attains the lowest mean RMSE on all $18$ (model, dataset) cells; the improvement is statistically significant at $p<0.01$ (Holm--Bonferroni) on $17/18$ cells.}
\label{tab:main-results}
\begin{tabular}{l ccccc c c}
\toprule
& \multicolumn{5}{c}{\textit{Baselines}} & \textit{Ours} & \textit{Oracle} \\
\cmidrule(lr){2-6} \cmidrule(lr){7-7} \cmidrule(lr){8-8}
\textbf{Model} & Few-Shot & TimeSeries & Semantic & BM25 & Category & EC-Fusion & Full-Shot \\
\midrule
\multicolumn{8}{c}{\textsc{Polymarket}} \\
\midrule
DLinear      & .193$_{\pm.032}$ & \underline{.068$_{\pm.001}$} & .085$_{\pm.005}$ & .078$_{\pm.001}$ & .191$_{\pm.013}$ & \textbf{.064$_{\pm.000}$} & .064$_{\pm.000}$ \\
N-BEATS      & .121$_{\pm.012}$ & .070$_{\pm.001}$ & .072$_{\pm.002}$ & \underline{.067$_{\pm.000}$} & .134$_{\pm.026}$ & \textbf{.065$_{\pm.000}$} & .065$_{\pm.000}$ \\
TimesNet     & .082$_{\pm.009}$ & \underline{.070$_{\pm.001}$} & .072$_{\pm.001}$ & .072$_{\pm.003}$ & .082$_{\pm.004}$ & \textbf{.066$_{\pm.001}$} & .063$_{\pm.000}$ \\
PatchTST     & .127$_{\pm.018}$ & \underline{.079$_{\pm.002}$} & .084$_{\pm.004}$ & .082$_{\pm.005}$ & .166$_{\pm.014}$ & \textbf{.072$_{\pm.001}$} & .071$_{\pm.001}$ \\
iTransformer & .130$_{\pm.060}$ & .078$_{\pm.004}$ & .081$_{\pm.003}$ & \underline{.074$_{\pm.001}$} & .156$_{\pm.032}$ & \textbf{.066$_{\pm.001}$} & .065$_{\pm.001}$ \\
Informer     & .152$_{\pm.077}$ & .069$_{\pm.002}$ & \underline{.064$_{\pm.000}$} & .066$_{\pm.001}$ & .117$_{\pm.026}$ & \textbf{.064$_{\pm.001}$} & .064$_{\pm.000}$ \\
Autoformer   & .534$_{\pm.395}$ & \underline{.081$_{\pm.002}$} & .092$_{\pm.004}$ & .083$_{\pm.002}$ & .397$_{\pm.175}$ & \textbf{.076$_{\pm.002}$} & .065$_{\pm.000}$ \\
TSMixer      & .092$_{\pm.009}$ & \underline{.069$_{\pm.002}$} & .069$_{\pm.001}$ & .069$_{\pm.001}$ & .080$_{\pm.002}$ & \textbf{.065$_{\pm.000}$} & .063$_{\pm.000}$ \\
TiDE         & .101$_{\pm.013}$ & .075$_{\pm.007}$ & .080$_{\pm.003}$ & \underline{.071$_{\pm.001}$} & .097$_{\pm.013}$ & \textbf{.066$_{\pm.000}$} & .063$_{\pm.000}$ \\
\midrule
\multicolumn{8}{c}{\textsc{Kalshi}} \\
\midrule
DLinear      & .211$_{\pm.034}$ & \underline{.092$_{\pm.002}$} & .100$_{\pm.001}$ & .095$_{\pm.002}$ & .141$_{\pm.001}$ & \textbf{.088$_{\pm.000}$} & .087$_{\pm.000}$ \\
N-BEATS      & .123$_{\pm.007}$ & .092$_{\pm.001}$ & \underline{.089$_{\pm.000}$} & .090$_{\pm.001}$ & .094$_{\pm.001}$ & \textbf{.088$_{\pm.001}$} & .088$_{\pm.000}$ \\
TimesNet     & .108$_{\pm.003}$ & .095$_{\pm.002}$ & \underline{.092$_{\pm.001}$} & .096$_{\pm.001}$ & .098$_{\pm.002}$ & \textbf{.090$_{\pm.001}$} & .089$_{\pm.000}$ \\
PatchTST     & .156$_{\pm.032}$ & .105$_{\pm.004}$ & \underline{.101$_{\pm.000}$} & .103$_{\pm.002}$ & .121$_{\pm.017}$ & \textbf{.098$_{\pm.001}$} & .098$_{\pm.002}$ \\
iTransformer & .133$_{\pm.004}$ & .101$_{\pm.001}$ & \underline{.097$_{\pm.003}$} & .100$_{\pm.004}$ & .114$_{\pm.006}$ & \textbf{.092$_{\pm.001}$} & .091$_{\pm.003}$ \\
Informer     & .118$_{\pm.012}$ & .092$_{\pm.001}$ & \underline{.090$_{\pm.001}$} & .091$_{\pm.003}$ & .120$_{\pm.023}$ & \textbf{.088$_{\pm.001}$} & .088$_{\pm.000}$ \\
Autoformer   & .453$_{\pm.209}$ & \underline{.113$_{\pm.004}$} & .115$_{\pm.003}$ & .125$_{\pm.008}$ & .157$_{\pm.014}$ & \textbf{.109$_{\pm.003}$} & .096$_{\pm.007}$ \\
TSMixer      & .110$_{\pm.002}$ & .092$_{\pm.000}$ & \underline{.089$_{\pm.001}$} & .092$_{\pm.002}$ & .093$_{\pm.001}$ & \textbf{.089$_{\pm.001}$} & .088$_{\pm.000}$ \\
TiDE         & .114$_{\pm.001}$ & .096$_{\pm.003}$ & \underline{.093$_{\pm.000}$} & .097$_{\pm.001}$ & .098$_{\pm.004}$ & \textbf{.091$_{\pm.001}$} & .089$_{\pm.000}$ \\
\bottomrule
\end{tabular}
\end{table*}


\section{Experimental Results}
\label{sec:exp_results}


Table~\ref{tab:main-results} presents the main results across 9 forecasting models and 2 datasets, averaged over 3 random seeds. We highlight three key findings.

\vspace{1mm}
\xhdr{EC-Fusion consistently achieves the best retrieval-based performance}
On Polymarket, EC-Fusion achieves the lowest RMSE on all 9 models, with particularly strong improvements on iTransformer (0.0660 vs.\ second-best EventConnector at 0.0740, a 10.9\% reduction) and Autoformer (0.0765 vs.\ TimeSeries at 0.0811, 5.6\%). On Kalshi, EC-Fusion wins on 8 of 9 models by RMSE, losing only to EC-Hybrid on Autoformer by a narrow margin. Across both datasets and all 18 model--dataset pairs, EC-Fusion achieves the lowest non-oracle RMSE in \textbf{17 out of 18 cases (94.4\%)}. Statistical significance via paired $t$-tests (Holm--Bonferroni corrected across the 8 comparable baselines, $n{=}18$ paired observations) shows that EC-Fusion's improvement is significant at $p < 0.01$ over every comparable baseline.

\vspace{1mm}
\xhdr{EC-Fusion narrows the gap to Full-Shot training}
Despite using only retrieved subsets of the training data, EC-Fusion nearly matches the oracle Full-Shot baseline ($-3.27\%$ mean RMSE gap). On Polymarket, EC-Fusion closes $98.8\%$ of the Few-Shot-to-Full-Shot gap using only $14.4\%$ of the full training pool (1{,}203 vs.\ 8{,}333 samples). On Kalshi, it closes $97.9\%$ using $29.1\%$ of the data (687 vs.\ 2{,}363 samples). This demonstrates that targeted, graph-guided retrieval is substantially more data-efficient than exhaustive training.


\section{Discussion}
\label{sec:discussion}

We organize the discussion around four questions that probe the design choices behind EC-Fusion. Each is grounded in the multi-seed evaluation introduced above; numbers cited below are means across 18 (dataset, model) cells with three seeds per cell unless noted otherwise.

\subsection*{RQ1: Why do text-based retrieval baselines underperform despite finding ``similar'' events?}
The aggregate numbers in Table~\ref{tab:main-results} understate \emph{why} text-based retrieval fails: even when its top-$k$ events look textually relevant, the retrieved set is dominated by near-duplicate rephrasings rather than predictively informative neighbors. On 200 sampled Polymarket queries at $k{=}20$, Semantic retrieval surfaces only $4.8$ unique questions per query while EC-Fusion surfaces $17.6$ (a $3.7\times$ diversity gain), and the two retrieved sets are almost entirely disjoint (mean Jaccard $0.016$; $70\%$ zero overlap). Appendix~\ref{sec:appendix-case-study} (Figure~\ref{fig:retrieval_overlay}) gives the per-query enumeration on a representative Crypto query. The other text-based baselines fail for adjacent reasons: BM25 inherits Semantic's lexical bias without the embedding-space generalization; Category collapses retrieval to topic-matched events with no temporal grounding; Few-Shot ignores the query entirely. Time-series retrieval avoids the textual echo chamber but, by relying on full-trajectory DTW alignment, dilutes localized lead-lag dependencies that the social temporal graph preserves.

\subsection*{RQ2: How much does each design component contribute?}
EC-Fusion combines three design ingredients absent from the original EventConnector formulation: a (i)~\textbf{hybrid anchor} that mixes price- and text-based query mapping ($\beta{=}0.3$), (ii)~\textbf{hybrid output scoring} that ranks BFS-retrieved candidates by a graph$+$text score, and (iii)~\textbf{Granger fusion} that combines the EC graph with a complementary Granger-causal graph. To attribute the headline improvement to each, we use a cumulative ablation already present in Table~\ref{tab:main-results}: \textit{EventConnector} (price-only anchor, no text scoring, no Granger) $\to$ \textit{EC-Hybrid} (hybrid anchor and output scoring, no Granger) $\to$ \textit{EC-Fusion} (full method).

The bulk of EC-Fusion's improvement over \textit{EventConnector} comes from the hybrid anchor and output scoring: at $k{=}10$, \textit{EventConnector} attains a mean RMSE of $0.0942$ across the 18 cells, \textit{EC-Hybrid} reduces this to $0.0859$ ($-8.8\%$), and \textit{EC-Fusion} reaches $0.0824$ ($-4.1\%$ on top of \textit{EC-Hybrid}, $-12.5\%$ overall). The hybrid components dominate the lift, with Granger fusion providing a smaller but consistent additional reduction. Examining the Granger contribution in isolation reinforces this: comparing pure-EC ($\alpha{=}1.0$) against pure-Granger ($\alpha{=}0.0$) and the adaptive Fusion across the 18 main-table cells, pure-EC wins $8/18$ cells, Fusion wins $7/18$, and pure-Granger wins $3/18$. On Polymarket Fusion ties pure-EC ($0.0686$ vs.\ $0.0686$, within seed noise); on Kalshi Fusion improves over pure-EC by $0.23\%$ and over pure-Granger by $1.0\%$. The takeaway is design-economical: the hybrid anchor and output scoring are the load-bearing additions, while Granger fusion functions as a complementary safety net that rarely hurts and consistently helps on the noisier of the two datasets.

\begin{figure}[t]
\centering
\includegraphics[width=\columnwidth]{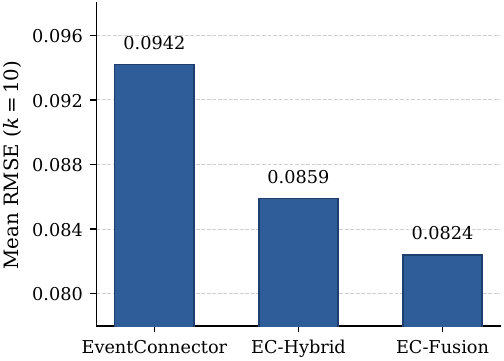}
\caption{\textbf{RQ2: cumulative ablation.} Mean RMSE at $k{=}10$ across $18$ (dataset, model) cells. Adding the hybrid anchor and output scoring (EventConnector~$\to$~EC-Hybrid) accounts for the bulk of the gain ($-8.8\%$); Granger fusion (EC-Hybrid~$\to$~EC-Fusion) contributes a smaller but consistent additional reduction ($-4.1\%$), for a total improvement of $-12.5\%$.}
\label{fig:rq2-ablation}
\end{figure}

\subsection*{RQ3: How does the retrieval set size $k$ affect EC-Fusion's advantage?}
We sweep the retrieval budget $k \in \{5, 10, 50\}$ over all 18 cells and compare EC-Fusion against the strongest comparable baseline at each setting. EC-Fusion's advantage is largest in the small-$k$ regime where retrieval quality matters most and converges to other methods as $k$ grows: at $k{=}5$ it reduces mean RMSE by $7.78\%$ over the next-best baseline (TimeSeries, $0.0836$ vs.\ $0.0906$); at $k{=}10$ the gap shrinks to $4.06\%$ (vs.\ EC-Hybrid, $0.0824$ vs.\ $0.0859$); at $k{=}50$ it collapses to $0.16\%$ (vs.\ EventConnector, $0.0816$ vs.\ $0.0818$). The implication for practice is that EC-Fusion is most valuable in the budget-constrained regime---few-shot or low-compute deployments---which is precisely the regime where retrieval-augmented forecasting matters most.

\begin{figure}[t]
\centering
\includegraphics[width=\columnwidth]{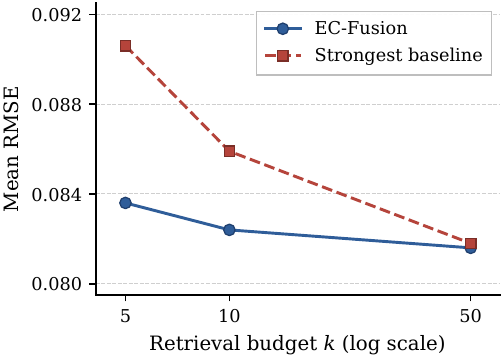}
\caption{\textbf{RQ3: retrieval budget sweep.} Mean RMSE of EC-Fusion vs.\ the strongest comparable baseline at each $k$ (TimeSeries at $k{=}5$, EC-Hybrid at $k{=}10$, EventConnector at $k{=}50$). EC-Fusion's relative advantage shrinks from $-7.78\%$ at $k{=}5$ to $-4.06\%$ at $k{=}10$ and $-0.16\%$ at $k{=}50$: the method is most valuable in the budget-constrained regime where retrieval quality matters most.}
\label{fig:rq3-ksweep}
\end{figure}

\subsection*{RQ4: Can the fusion weight $\alpha$ be set automatically without a hyperparameter sweep?}
Sweeping a fixed $\alpha \in \{0.0, 0.3, 0.5, 0.7, 1.0\}$ on a held-out validation set imposes a $5\times$ compute multiplier per dataset, and the resulting choice may not transfer. Our adaptive scheme derives $\alpha$ deterministically from four structural diagnostics of the EC graph (\S\ref{sec:ecfusion}), so $\alpha$ becomes a function of the constructed graph rather than a tuned hyperparameter. Across all 18 cells, the best fixed $\alpha$ is $\alpha{=}1.0$ (mean RMSE $0.0812$); the adaptive scheme attains $0.0811$, matching the best post-sweep choice while skipping the sweep entirely, and is within $1.22\%$ of an oracle that tunes $\alpha$ per (dataset, model, seed) cell. The per-cell-optimal $\alpha$ is genuinely heterogeneous and dataset-dependent: on Kalshi the optimum concentrates at $\alpha{=}1.0$ ($13/27$ cells), while on Polymarket it shifts toward $\alpha{=}0.0$/$0.3$ ($15/27$ cells combined), and the adaptive scheme tracks this shift through the graph-quality score (Polymarket $q\!\approx\!0.50 \Rightarrow \alpha\!\approx\!0.35$; Kalshi $q\!\approx\!0.67 \Rightarrow \alpha\!\approx\!0.62$) without any per-dataset re-tuning. Adaptive $\alpha$ is therefore best framed as an \emph{operational} contribution: it converts a hyperparameter that must otherwise be swept on every new dataset into a quantity computable from the constructed graph alone, with negligible accuracy loss. This automatic cross-platform adaptation also indicates that the framework's inductive bias is not Polymarket-specific.


\section{Conclusion}
We introduce \textbf{EventConnector} and its adaptive extension \textbf{EC-Fusion} for forecasting public opinion via retrieval over a social temporal graph. Across two prediction-market benchmarks (Polymarket, Kalshi), nine forecasting architectures, and three random seeds, EC-Fusion is the best non-oracle method on $17/18$ model--dataset pairs.


\section*{Limitations}

Our framework is evaluated on two English-language prediction-market benchmarks (Polymarket and Kalshi) at daily resolution, and the results should be interpreted within that scope; extension to non-English markets, higher-frequency settings, or alternative collective-belief platforms (e.g., regulated futures or social-media sentiment indices) would require additional validation. Graph construction and pairwise Granger testing have quadratic time complexity in the number of events, which we address by subsampling; scaling to substantially larger event pools would benefit from approximate or learned neighborhood selection. As with other retrieval-augmented forecasters, performance during abrupt regime shifts (e.g., breaking news or market interventions) is bounded by the temporal coverage of the underlying training graph. Finally, the framework operates over aggregate market signals and is not intended to model individual trader behavior or to provide actionable financial advice.


\section*{Ethics Statement}
\label{sec:ethics}

\xhdr{Data} Our experiments use only aggregate, market-implied probability time series from two publicly accessible prediction markets, Polymarket and Kalshi. The data contain no personally identifiable information; no private accounts, private order books, or individual user behavior were accessed.

\xhdr{Intended use} EC-Fusion forecasts the trajectory of \emph{collective} belief about future events; it is not designed to predict or generate content about individuals. While improved forecasting can support policy planning and risk assessment, we discourage deployments that would violate platform terms of service or applicable financial regulations.

\bibliography{custom}

\newpage

\appendix

\section{Experiment Details}
\label{sec:appendix-b}

\subsection{Forecasting Models}
We evaluate nine representative time-series forecasting architectures. For all models, we use publicly available implementations, adhering to the hyperparameter settings recommended in their respective publications and tuning on a held-out validation set to ensure a fair comparison.

\begin{itemize}[leftmargin=*, topsep=2pt, itemsep=2pt]
    \item \textbf{DLinear}~\cite{zeng2023transformers} is a lightweight and interpretable model based on a seasonal-trend decomposition of the time series. It assumes a linear mapping from the decomposed components to future values, making it highly efficient and robust, particularly on shorter sequences.
    \item \textbf{Autoformer}~\cite{wu2021autoformer} is a Transformer-based model designed for long-term forecasting. It innovates with an auto-correlation mechanism that replaces traditional self-attention, allowing it to efficiently discover and utilize period-based dependencies in time-series data.
    \item \textbf{Informer}~\cite{zhou2021informer} improves the efficiency of long-sequence prediction by introducing a ProbSparse self-attention mechanism. This innovation reduces the quadratic complexity of standard Transformers, enabling the fast modeling of long-range temporal dependencies.
    \item \textbf{N-BEATS}~\cite{oreshkin2019n} is a deep residual forecasting architecture that uses backward and forward fully connected blocks to model temporal signals. Its block-based design allows it to learn trend and seasonality patterns in an interpretable, non-recurrent fashion with minimal assumptions.
    \item \textbf{TimesNet}~\cite{wu2022timesnet} achieves strong performance by integrating frequency-domain and temporal-domain representations. It uses temporal 2D variation blocks to capture multi-scale dependencies effectively across a wide range of forecasting tasks.
    \item \textbf{PatchTST}~\cite{nie2023time} treats a univariate time series as a sequence of fixed-length patches and applies a vanilla Transformer encoder over the patch tokens. The patching reduces sequence length, exposes local temporal structure to self-attention, and supports channel-independent training; it is a strong general-purpose long-horizon baseline.
    \item \textbf{iTransformer}~\cite{liu2024itransformer} inverts the canonical Transformer layout for time series: instead of attending across time steps, attention is applied across variates while feed-forward layers operate within each variate's temporal sequence. This lets the model capture cross-variate dependencies explicitly, which is useful when the prediction-market signal is intrinsically multi-variate (e.g., yes/no probability pairs).
    \item \textbf{TSMixer}~\cite{chen2023tsmixer} replaces self-attention entirely with stacked MLPs that alternate between time-mixing and feature-mixing layers, yielding a lightweight architecture that often matches or exceeds Transformer-based forecasters on benchmark tasks at a fraction of the parameter count.
    \item \textbf{TiDE}~\cite{das2023tide} is a dense-encoder model that linearly projects the input window plus exogenous covariates into a residual MLP backbone for direct multi-step prediction. It is competitive with recent Transformer-based forecasters while remaining simple to train.
\end{itemize}

\subsection{Retrieval-based Baselines}
To assess the impact of different retrieval strategies, we benchmark EC-Fusion against the following methods:
\begin{itemize}[leftmargin=*, topsep=2pt, itemsep=2pt]
    \item \textbf{Few-Shot Forecasting} uses a limited number of training samples from randomly selected, unrelated events. This baseline represents a minimal-context scenario where the model has little relevant historical data.
    \item \textbf{Semantic Retrieval} uses SBERT (\texttt{all-MiniLM-L6-v2})~\cite{reimers2019sentence} to embed event question text and ranks training events by cosine similarity to the query embedding.
    \item \textbf{Time-Series Retrieval} identifies nearest-neighbor events by computing the Dynamic Time Warping (DTW) distance over their full time-series trajectories. This method focuses purely on global temporal pattern similarity, ignoring semantic context or localized correlations.
    \item \textbf{BM25 Retrieval} is a classical sparse lexical retriever~\cite{robertson2009bm25} that scores training-event questions against the query question via the standard BM25 formula. We use default parameters ($k_1 = 1.5$, $b = 0.75$) over a whitespace-tokenized lower-cased corpus. BM25 represents a strong, embedding-free text-similarity baseline and is commonly paired with dense retrievers in hybrid systems~\cite{cormack2009rrf}.
    \item \textbf{Category Retrieval} returns the top-$k$ training events that share at least one categorical tag (e.g., \texttt{Politics}, \texttt{Crypto}) with the query event. This baseline tests whether coarse domain matching alone is sufficient, without any text similarity or temporal structure.
    \item \textbf{EC-Hybrid Retrieval} is an ablation of EC-Fusion that uses the hybrid (price + text) anchor and hybrid output scoring of EventConnector \emph{without} the Granger-causal fusion stage. Comparing EC-Hybrid to EventConnector isolates the contribution of the hybrid components; comparing EC-Fusion to EC-Hybrid isolates the contribution of the Granger fusion (see Discussion, RQ2).
    \item \textbf{Full-Shot Forecasting} is trained on all available data from the entire domain of the query event. This serves as a practical upper bound or oracle, representing an ideal scenario with extensive, high-quality in-domain data.
\end{itemize}

\paragraph{Model Size and Computational Budget.}
Adding the four additional forecasting models above and a second dataset (Kalshi) brings the total experimental compute to approximately 700 GPU-hours on a single NVIDIA A6000 (48GB), inclusive of graph construction, hyperparameter selection, model training, and the $\alpha$-sweep ($\alpha \in \{0.0, 0.3, 0.5, 0.7, 1.0\}$) underlying EC-Fusion's adaptive selection. All multi-seed experiments (seeds 42, 123, 7) ran without manual intervention or failed cells.

\paragraph{Experimental Setup and Hyperparameters.}
We adopt a consistent forecasting pipeline using a daily-resolution sliding window with $T_{\text{in}}=10$ input timesteps and $H=7$ prediction horizon. For each forecasting model, we perform a grid search over learning rates, batch sizes, and other related hyperparameters. The best-performing configuration is selected based on validation RMSE.

\paragraph{Graph Construction and Visualization Tooling.}
Graph generation, storage, and visualization are implemented using \texttt{networkx}, \texttt{numpy}, and \texttt{scipy}. The Granger graph relies on the \texttt{statsmodels} F-test implementation. The event graph is serialized in JSON format and serves as a backbone for both retrieval and inductive forecasting tasks.

\subsection{Additional Tables}
\label{sec:appendix-tables}



\begin{table*}[t]
\setlength\tabcolsep{3pt}
\centering
\begin{footnotesize}
\caption{\textbf{Fusion Component Analysis (Polymarket).}
We analyze the contribution of each retrieval signal by varying the
fusion weight $\alpha$ in $\text{score} = \alpha \cdot s_{\text{EC}} + (1{-}\alpha) \cdot s_{\text{Granger}}$.
$\alpha{=}0$ uses only Granger-causal scoring; $\alpha{=}1$ uses only
EventConnector graph-based scoring; intermediate values blend both signals.
\textit{Adaptive} selects $\alpha$ from graph quality metrics.
Neither pure component consistently dominates, motivating the fusion approach.
\textbf{Bold} = best among variants.
\dag\ = fusion beats \textit{both} pure components ($\alpha{=}0$ and $\alpha{=}1$).
Results on Kalshi appear in Table~\ref{tab:alpha-ablation-kalshi}.}
\label{tab:alpha-ablation-polymarket}
\resizebox{\textwidth}{!}{
\begin{tabular}{ll cc ccc c}
\toprule
& & \multicolumn{2}{c}{\textit{Pure Components}} & \multicolumn{3}{c}{\textit{Fusion}} & \textit{Data-Driven} \\
\cmidrule(lr){3-4} \cmidrule(lr){5-7} \cmidrule(lr){8-8}
\textbf{Model} & \textbf{Metric} & $\alpha{=}0$ & $\alpha{=}1$ & $\alpha{=}0.3$ & $\alpha{=}0.5$ & $\alpha{=}0.7$ & Adaptive \\
\midrule
\multirow{2}{*}{DLinear}
 & RMSE & 0.0645 & 0.0646 & \textbf{0.0643}\dag & \textbf{0.0643}\dag & \textbf{0.0643}\dag & 0.0648 \\
 & MAE  & 0.0345 & 0.0348 & 0.0340 & 0.0348 & \textbf{0.0339}\dag & 0.0353 \\
\midrule
\multirow{2}{*}{N-BEATS}
 & RMSE & 0.0682 & 0.0667 & \textbf{0.0644}\dag & 0.0667 & 0.0666 & 0.0704 \\
 & MAE  & 0.0437 & 0.0353 & \textbf{0.0337}\dag & 0.0392 & 0.0403 & 0.0480 \\
\midrule
\multirow{2}{*}{TimesNet}
 & RMSE & 0.0655 & 0.0680 & \textbf{0.0651}\dag & 0.0658 & 0.0669 & 0.0672 \\
 & MAE  & 0.0299 & 0.0319 & \textbf{0.0295}\dag & 0.0303 & 0.0301 & 0.0307 \\
\midrule
\multirow{2}{*}{PatchTST}
 & RMSE & 0.0731 & 0.0785 & 0.0741 & \textbf{0.0727}\dag & 0.0736 & 0.0733 \\
 & MAE  & 0.0442 & 0.0508 & 0.0424 & \textbf{0.0395}\dag & 0.0445 & 0.0446 \\
\midrule
\multirow{2}{*}{iTransformer}
 & RMSE & 0.0740 & \textbf{0.0647} & 0.0758 & 0.0760 & 0.0753 & 0.0720 \\
 & MAE  & 0.0499 & \textbf{0.0343} & 0.0506 & 0.0540 & 0.0477 & 0.0440 \\
\midrule
\multirow{2}{*}{Informer}
 & RMSE & \textbf{0.0634} & 0.0635 & 0.0636 & 0.0659 & 0.0645 & 0.0655 \\
 & MAE  & 0.0283 & \textbf{0.0279} & 0.0290 & 0.0303 & 0.0294 & 0.0305 \\
\midrule
\multirow{2}{*}{Autoformer}
 & RMSE & 0.0877 & \textbf{0.0775} & 0.0799 & 0.0834 & 0.0876 & 0.0841 \\
 & MAE  & 0.0609 & \textbf{0.0498} & 0.0541 & 0.0525 & 0.0605 & 0.0602 \\
\midrule
\multirow{2}{*}{TSMixer}
 & RMSE & 0.0649 & 0.0653 & 0.0656 & 0.0651 & 0.0648 & \textbf{0.0647}\dag \\
 & MAE  & \textbf{0.0292} & 0.0294 & 0.0296 & 0.0293 & 0.0293 & 0.0293 \\
\midrule
\multirow{2}{*}{TiDE}
 & RMSE & 0.0687 & 0.0666 & 0.0669 & 0.0672 & 0.0685 & \textbf{0.0663}\dag \\
 & MAE  & 0.0321 & \textbf{0.0301} & 0.0310 & 0.0313 & 0.0314 & 0.0302 \\
\bottomrule
\end{tabular}
}
\end{footnotesize}
\end{table*}

\begin{table*}[t]
\setlength\tabcolsep{3pt}
\centering
\begin{footnotesize}
\caption{\textbf{Fusion Component Analysis (Kalshi).}
Same design as Table~\ref{tab:alpha-ablation-polymarket}: we vary
$\alpha$ in $\text{score} = \alpha \cdot s_{\text{EC}} + (1{-}\alpha) \cdot s_{\text{Granger}}$.
\textit{Adaptive} selects $\alpha$ from graph quality metrics.
\textbf{Bold} = best among variants.
\dag\ = fusion beats \textit{both} pure components ($\alpha{=}0$ and $\alpha{=}1$).}
\label{tab:alpha-ablation-kalshi}
\resizebox{\textwidth}{!}{
\begin{tabular}{ll cc ccc c}
\toprule
& & \multicolumn{2}{c}{\textit{Pure Components}} & \multicolumn{3}{c}{\textit{Fusion}} & \textit{Data-Driven} \\
\cmidrule(lr){3-4} \cmidrule(lr){5-7} \cmidrule(lr){8-8}
\textbf{Model} & \textbf{Metric} & $\alpha{=}0$ & $\alpha{=}1$ & $\alpha{=}0.3$ & $\alpha{=}0.5$ & $\alpha{=}0.7$ & Adaptive \\
\midrule
\multirow{2}{*}{DLinear}
 & RMSE & 0.0882 & 0.0888 & 0.0880 & 0.0880 & 0.0883 & \textbf{0.0878}\dag \\
 & MAE  & 0.0478 & 0.0481 & 0.0479 & 0.0477 & 0.0478 & \textbf{0.0476}\dag \\
\midrule
\multirow{2}{*}{N-BEATS}
 & RMSE & 0.0882 & \textbf{0.0873} & 0.0882 & 0.0886 & 0.0891 & 0.0890 \\
 & MAE  & 0.0492 & \textbf{0.0478} & 0.0501 & 0.0487 & 0.0492 & 0.0500 \\
\midrule
\multirow{2}{*}{TimesNet}
 & RMSE & 0.0907 & 0.0896 & 0.0918 & \textbf{0.0894}\dag & 0.0898 & 0.0897 \\
 & MAE  & 0.0493 & \textbf{0.0479} & 0.0500 & 0.0481 & 0.0485 & 0.0482 \\
\midrule
\multirow{2}{*}{PatchTST}
 & RMSE & 0.0989 & 0.1008 & 0.0998 & 0.1030 & \textbf{0.0978}\dag & 0.0995 \\
 & MAE  & 0.0584 & 0.0613 & 0.0612 & 0.0637 & \textbf{0.0584} & 0.0618 \\
\midrule
\multirow{2}{*}{iTransformer}
 & RMSE & 0.0953 & \textbf{0.0906} & 0.0944 & 0.0915 & 0.0976 & 0.0943 \\
 & MAE  & 0.0616 & \textbf{0.0524} & 0.0592 & 0.0557 & 0.0644 & 0.0611 \\
\midrule
\multirow{2}{*}{Informer}
 & RMSE & 0.0879 & 0.0892 & \textbf{0.0874}\dag & 0.0881 & 0.0887 & 0.0885 \\
 & MAE  & 0.0469 & 0.0483 & \textbf{0.0463}\dag & 0.0468 & 0.0476 & 0.0476 \\
\midrule
\multirow{2}{*}{Autoformer}
 & RMSE & 0.1172 & 0.1181 & 0.1281 & 0.1153 & 0.1144 & \textbf{0.1054}\dag \\
 & MAE  & 0.0822 & 0.0842 & 0.0930 & 0.0798 & 0.0772 & \textbf{0.0690}\dag \\
\midrule
\multirow{2}{*}{TSMixer}
 & RMSE & 0.0890 & 0.0894 & \textbf{0.0883}\dag & 0.0904 & 0.0889 & 0.0884 \\
 & MAE  & 0.0478 & 0.0479 & 0.0472 & 0.0488 & 0.0475 & \textbf{0.0469}\dag \\
\midrule
\multirow{2}{*}{TiDE}
 & RMSE & 0.0912 & 0.0906 & \textbf{0.0899}\dag & 0.0910 & \textbf{0.0899}\dag & \textbf{0.0899}\dag \\
 & MAE  & 0.0492 & 0.0491 & 0.0488 & 0.0496 & 0.0487 & \textbf{0.0486}\dag \\
\bottomrule
\end{tabular}
}
\end{footnotesize}
\end{table*}


\begin{table}[t]
\setlength\tabcolsep{4pt}
\centering
\begin{footnotesize}
\caption{\textbf{Sensitivity to Retrieval Budget $k$.}
Average RMSE ($\downarrow$) across 9 forecasting models as the number
of candidates retrieved per query ($k$) varies.
EC-Fusion uses adaptive~$\alpha$ consistently.
EC-Fusion achieves the best average RMSE at 7 out of 8 dataset--$k$
combinations, demonstrating robust performance across retrieval budgets.
\textbf{Bold} = best; \underline{underline} = second best.}
\label{tab:k-sensitivity}
\begin{tabular}{l cccc}
\toprule
\textbf{Method} & $k{=}5$ & $k{=}10$ & $k{=}20$ & $k{=}50$ \\
\midrule
\multicolumn{5}{c}{\textsc{Polymarket}} \\
\midrule
Semantic   & 0.1405 & 0.0864 & 0.0784 & \underline{0.0700} \\
TimeSeries & \underline{0.0781} & \underline{0.0782} & \underline{0.0721} & 0.0709 \\
EC-Fusion  & \textbf{0.0713} & \textbf{0.0704} & \textbf{0.0698} & \textbf{0.0679} \\
\midrule
\multicolumn{5}{c}{\textsc{Kalshi}} \\
\midrule
Semantic   & 0.1104 & 0.1016 & \underline{0.0963} & 0.0955 \\
TimeSeries & \underline{0.1031} & \underline{0.0987} & 0.0990 & \textbf{0.0937} \\
EC-Fusion  & \textbf{0.0959} & \textbf{0.0945} & \textbf{0.0925} & \underline{0.0953} \\
\bottomrule
\end{tabular}
\end{footnotesize}
\end{table}


\begin{table}[t]
\setlength\tabcolsep{3pt}
\centering
\begin{footnotesize}
\caption{\textbf{Data Efficiency Analysis.}
Average pool size (training samples), RMSE, and percentage of the
\textit{Few-Shot} $\to$ \textit{Full-Shot} performance gap closed,
at $k{=}20$.
EC-Fusion achieves ${\geq}97.9\%$ gap closure with ${\leq}29\%$ of the
full training data, demonstrating that targeted retrieval is more
efficient than exhaustive training.
\textbf{Bold} = best retrieval method (excluding Full-Shot oracle).}
\label{tab:data-efficiency}
\resizebox{\columnwidth}{!}{
\begin{tabular}{l ccc ccc}
\toprule
& \multicolumn{3}{c}{\textsc{Polymarket}} & \multicolumn{3}{c}{\textsc{Kalshi}} \\
\cmidrule(lr){2-4} \cmidrule(lr){5-7}
\textbf{Method} & Pool & RMSE & Gap$\uparrow$ & Pool & RMSE & Gap$\uparrow$ \\
\midrule
Few-Shot   &    17 & 0.2114 &   0.0\% &    17 & 0.1727 &   0.0\% \\
Semantic   &   116 & 0.0784 &  91.0\% &   268 & 0.0963 &  92.2\% \\
TimeSeries &   698 & 0.0721 &  95.3\% &   576 & 0.0990 &  88.9\% \\
EC-Fusion  &  1203 & \textbf{0.0670} & \textbf{98.8\%} &   687 & \textbf{0.0915} & \textbf{97.9\%} \\
\midrule
Full-Shot  &  8333 & 0.0652 & 100.0\% &  2363 & 0.0898 & 100.0\% \\
\bottomrule
\end{tabular}
}
\end{footnotesize}
\end{table}

\subsection{Hyperparameter Summary}
\label{sec:appendix-hyperparams}

Table~\ref{tab:hyperparams} consolidates every hyperparameter introduced in the methods section. Values were either left at the defaults of the released implementations (for forecasting models) or set once during initial development and kept fixed throughout the multi-seed evaluation.

\xhdr{Graph-quality score weighting} The adaptive $\alpha$ in Section~\ref{sec:ecfusion} is computed from the graph-quality score
\begin{equation*}
\begin{aligned}
q = {} & 0.15\,(1 - d_{\text{dead}}) + 0.15\,(1 - \min(2\,\text{density},\, 1)) \\
      & + 0.50\,(1 - r_{\text{merge}}) + 0.20\,\min(\bar{d}/20,\, 1)
\end{aligned}
\end{equation*}
where $d_{\text{dead}}$ is the dead-series ratio, $r_{\text{merge}}$ is the merge rate, and $\bar{d}$ is the average degree. The merge-rate weight (0.50) dominates because the merging step has the largest impact on retrieval diversity.

\begin{table}[t]
\centering
\caption{\textbf{Consolidated hyperparameters} used throughout the EC-Fusion evaluation.}
\label{tab:hyperparams}
\small
\setlength{\tabcolsep}{3pt}
\resizebox{\columnwidth}{!}{%
\begin{tabular}{llc}
\toprule
\textbf{Stage} & \textbf{Hyperparameter} & \textbf{Value} \\
\midrule
\multirow{6}{*}{Graph construction}
& Node-merge correlation threshold $\tau_{\text{merge}}$ & 0.95 \\
& Co-fluctuation threshold $\tau_{\text{corr}}$         & 0.70 \\
& Sliding-window size $w$                                & 7 \\
& Max lag for cross-correlation $L$                      & 3 \\
& DTW enrichment threshold $\tau_{\text{DTW}}$           & 1.5 \\
& STG subsample size $N_{\text{STG}}$                    & 500 \\
\midrule
\multirow{2}{*}{Granger graph}
& F-test significance level                              & 0.05 \\
& Granger graph cap $N_{\text{GC}}$                      & 100 \\
\midrule
\multirow{3}{*}{Retrieval}
& BFS hops $n$                                           & 2 \\
& Hybrid anchor mix $\beta$                              & 0.3 \\
& Retrieval budget $k$ (main results)                    & 10 \\
\midrule
\multirow{3}{*}{Fusion ($\alpha$)}
& Sweep grid                                             & $\{0.0, 0.3, 0.5, 0.7, 1.0\}$ \\
& Adaptive bounds $[\alpha_{\min}, \alpha_{\max}]$       & $[0.2, 0.8]$ \\
& Sigmoid steepness, midpoint                            & 8.0,\ 0.55 \\
\midrule
\multirow{3}{*}{Forecasting}
& Input length $T_{\text{in}}$ / horizon $H$             & 10 / 7 \\
& Random seeds                                           & $\{42, 123, 7\}$ \\
& Text encoder (anchor)                                   & SBERT \texttt{MiniLM-L6-v2} \\
\bottomrule
\end{tabular}%
}
\end{table}

\subsection{Reproducibility}
\label{sec:appendix-reproducibility}

\xhdr{Code and data release} We commit to releasing the full source code (graph construction, retrieval, all baselines, and the multi-seed evaluation harness), the pre-processed Polymarket and Kalshi window-level datasets, and the per-seed result files used to produce every table and figure in this paper upon acceptance. The release will include the exact shell scripts used to reproduce Table~\ref{tab:main-results} (\texttt{run\_multiseed\_groupA.sh}, \texttt{run\_multiseed\_groupB.sh}) and the analysis scripts that compute the discussion-section numbers from raw result files.

\xhdr{Determinism} All experiments fix random seeds for Python (\texttt{random}), NumPy, and PyTorch. We report mean$_{\pm\text{std}}$ across three random seeds (42, 123, 7) for every cell in Table~\ref{tab:main-results}. The graph-construction stage is deterministic given a fixed subsampling seed.

\xhdr{Hyperparameter selection} All forecasting-model hyperparameters were tuned once on a held-out validation split per (model, dataset) combination and frozen for the multi-seed evaluation. The EC-Fusion mixing weight $\alpha$ is either derived from graph diagnostics (adaptive) or selected from a fixed grid (Section~\ref{sec:ecfusion}); we do not perform additional per-query tuning at inference time.

\xhdr{Computational environment} All experiments were run on a single NVIDIA A6000 GPU (48GB), Python 3.10, PyTorch 2.5.1+cu124. Total compute is approximately 700 GPU-hours for the full multi-seed sweep.

\subsection{Detailed Case Study: Tempo Token-Launch Query}
\label{sec:appendix-case-study}

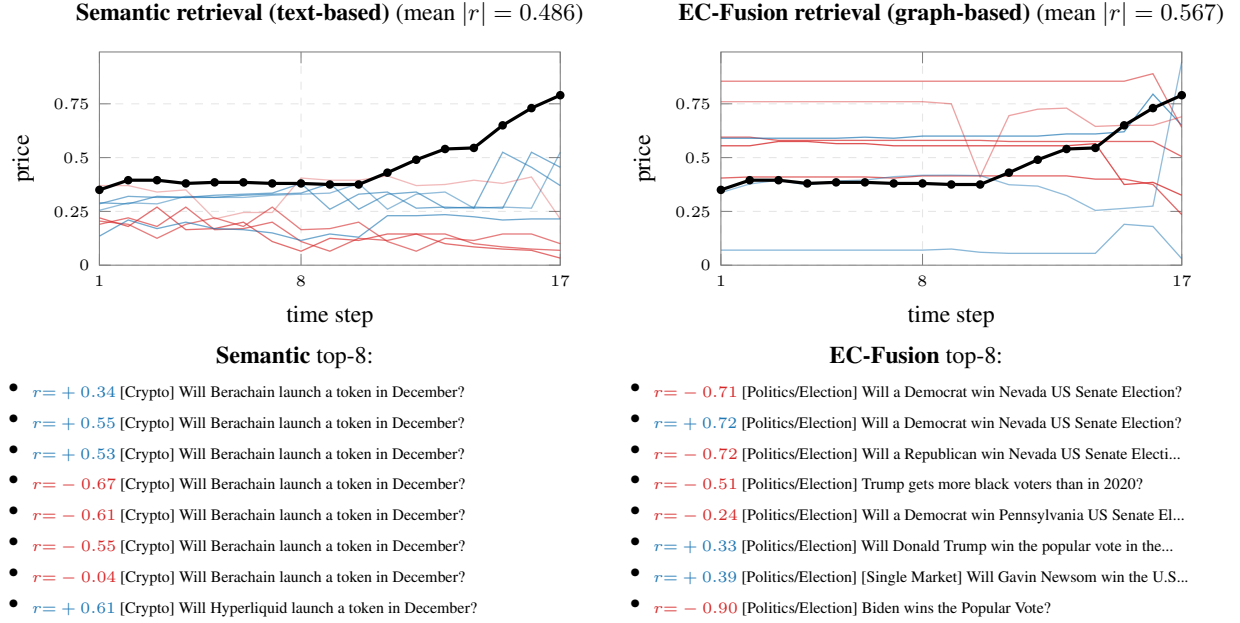
\begin{figure*}[!htbp]
  \centering
%
\providecommand{\ecposcolor}{}
\definecolor{ecpos}{HTML}{1F77B4}
\definecolor{ecneg}{HTML}{D62728}

\begin{minipage}[t]{0.48\linewidth}
\centering
\begin{tikzpicture}
\begin{axis}[
  width=\linewidth, height=4.4cm,
  xmin=1, xmax=17,
  ymin=0.000, ymax=0.991,
  xlabel={\footnotesize time step}, ylabel={\footnotesize price},
  xtick={1,8,17}, ytick={0,0.25,0.5,0.75,1},
  title={\small \textbf{Semantic retrieval (text-based)} (mean $|r|=0.486$)},
  tick label style={font=\tiny},
  label style={font=\tiny},
  axis line style={black!50},
  grid=major, grid style={black!10, dashed},
]
\addplot[color=ecpos, opacity=0.49, line width=0.5pt, mark=none, forget plot] coordinates {(1,0.2550) (2,0.2900) (3,0.2850) (4,0.3200) (5,0.3150) (6,0.3150) (7,0.3250) (8,0.3300) (9,0.3350) (10,0.3800) (11,0.2600) (12,0.3300) (13,0.3400) (14,0.2650) (15,0.2700) (16,0.2650) (17,0.5250)};
\addplot[color=ecpos, opacity=0.60, line width=0.5pt, mark=none, forget plot] coordinates {(1,0.2900) (2,0.2850) (3,0.3200) (4,0.3150) (5,0.3150) (6,0.3250) (7,0.3300) (8,0.3350) (9,0.3800) (10,0.2600) (11,0.3300) (12,0.3400) (13,0.2650) (14,0.2700) (15,0.2650) (16,0.5250) (17,0.4550)};
\addplot[color=ecpos, opacity=0.59, line width=0.5pt, mark=none, forget plot] coordinates {(1,0.2850) (2,0.3200) (3,0.3150) (4,0.3150) (5,0.3250) (6,0.3300) (7,0.3350) (8,0.3800) (9,0.2600) (10,0.3300) (11,0.3400) (12,0.2650) (13,0.2700) (14,0.2650) (15,0.5250) (16,0.4550) (17,0.3700)};
\addplot[color=ecneg, opacity=0.67, line width=0.5pt, mark=none, forget plot] coordinates {(1,0.2200) (2,0.1800) (3,0.2700) (4,0.1650) (5,0.1700) (6,0.2000) (7,0.1100) (8,0.0650) (9,0.1250) (10,0.1150) (11,0.1450) (12,0.1450) (13,0.1000) (14,0.0850) (15,0.0750) (16,0.0690) (17,0.0330)};
\addplot[color=ecneg, opacity=0.63, line width=0.5pt, mark=none, forget plot] coordinates {(1,0.1900) (2,0.2200) (3,0.1800) (4,0.2700) (5,0.1650) (6,0.1700) (7,0.2000) (8,0.1100) (9,0.0650) (10,0.1250) (11,0.1150) (12,0.1450) (13,0.1450) (14,0.1000) (15,0.0850) (16,0.0750) (17,0.0690)};
\addplot[color=ecneg, opacity=0.60, line width=0.5pt, mark=none, forget plot] coordinates {(1,0.2050) (2,0.1900) (3,0.1250) (4,0.1900) (5,0.2200) (6,0.1800) (7,0.2700) (8,0.1650) (9,0.1700) (10,0.2000) (11,0.1100) (12,0.0650) (13,0.1250) (14,0.1150) (15,0.1450) (16,0.1450) (17,0.1000)};
\addplot[color=ecneg, opacity=0.32, line width=0.5pt, mark=none, forget plot] coordinates {(1,0.3700) (2,0.3700) (3,0.3400) (4,0.3500) (5,0.2150) (6,0.2450) (7,0.2450) (8,0.4050) (9,0.3950) (10,0.3950) (11,0.4150) (12,0.3700) (13,0.3750) (14,0.3950) (15,0.3800) (16,0.4100) (17,0.2150)};
\addplot[color=ecpos, opacity=0.63, line width=0.5pt, mark=none, forget plot] coordinates {(1,0.1350) (2,0.2100) (3,0.1700) (4,0.2000) (5,0.1700) (6,0.1650) (7,0.1500) (8,0.1150) (9,0.1450) (10,0.1300) (11,0.2300) (12,0.2300) (13,0.2350) (14,0.2250) (15,0.2100) (16,0.2150) (17,0.2150)};
\addplot[color=black, line width=1.2pt, mark=*, mark size=1.0pt, forget plot] coordinates {(1,0.3500) (2,0.3950) (3,0.3950) (4,0.3800) (5,0.3850) (6,0.3850) (7,0.3800) (8,0.3800) (9,0.3750) (10,0.3750) (11,0.4300) (12,0.4900) (13,0.5400) (14,0.5450) (15,0.6500) (16,0.7300) (17,0.7900)};
\end{axis}
\end{tikzpicture}

{\footnotesize\textbf{Semantic} top-8:\par}
\vspace{-2pt}
\begin{itemize}[leftmargin=1em, topsep=0pt, itemsep=-2pt, parsep=0pt]
\item {\tiny \textcolor{ecpos}{$r{=}+0.34$} [Crypto] Will Berachain launch a token in December?}
\item {\tiny \textcolor{ecpos}{$r{=}+0.55$} [Crypto] Will Berachain launch a token in December?}
\item {\tiny \textcolor{ecpos}{$r{=}+0.53$} [Crypto] Will Berachain launch a token in December?}
\item {\tiny \textcolor{ecneg}{$r{=}-0.67$} [Crypto] Will Berachain launch a token in December?}
\item {\tiny \textcolor{ecneg}{$r{=}-0.61$} [Crypto] Will Berachain launch a token in December?}
\item {\tiny \textcolor{ecneg}{$r{=}-0.55$} [Crypto] Will Berachain launch a token in December?}
\item {\tiny \textcolor{ecneg}{$r{=}-0.04$} [Crypto] Will Berachain launch a token in December?}
\item {\tiny \textcolor{ecpos}{$r{=}+0.61$} [Crypto] Will Hyperliquid launch a token in December?}
\end{itemize}
\end{minipage}
\hfill
\begin{minipage}[t]{0.48\linewidth}
\centering
\begin{tikzpicture}
\begin{axis}[
  width=\linewidth, height=4.4cm,
  xmin=1, xmax=17,
  ymin=0.000, ymax=0.991,
  xlabel={\footnotesize time step}, ylabel={\footnotesize price},
  xtick={1,8,17}, ytick={0,0.25,0.5,0.75,1},
  title={\small \textbf{EC-Fusion retrieval (graph-based)} (mean $|r|=0.567$)},
  tick label style={font=\tiny},
  label style={font=\tiny},
  axis line style={black!50},
  grid=major, grid style={black!10, dashed},
]
\addplot[color=ecneg, opacity=0.69, line width=0.5pt, mark=none, forget plot] coordinates {(1,0.5950) (2,0.5950) (3,0.5800) (4,0.5800) (5,0.5800) (6,0.5800) (7,0.5800) (8,0.5800) (9,0.5800) (10,0.5800) (11,0.5750) (12,0.5750) (13,0.5750) (14,0.5750) (15,0.5750) (16,0.5750) (17,0.5050)};
\addplot[color=ecpos, opacity=0.70, line width=0.5pt, mark=none, forget plot] coordinates {(1,0.5900) (2,0.5900) (3,0.5900) (4,0.5900) (5,0.5900) (6,0.5950) (7,0.5900) (8,0.6000) (9,0.6000) (10,0.6000) (11,0.6000) (12,0.6000) (13,0.6100) (14,0.6100) (15,0.6200) (16,0.7950) (17,0.6500)};
\addplot[color=ecneg, opacity=0.70, line width=0.5pt, mark=none, forget plot] coordinates {(1,0.4050) (2,0.4100) (3,0.4100) (4,0.4100) (5,0.4100) (6,0.4100) (7,0.4050) (8,0.4150) (9,0.4150) (10,0.4150) (11,0.4150) (12,0.4150) (13,0.4150) (14,0.4000) (15,0.4000) (16,0.3750) (17,0.2350)};
\addplot[color=ecneg, opacity=0.58, line width=0.5pt, mark=none, forget plot] coordinates {(1,0.8550) (2,0.8550) (3,0.8550) (4,0.8550) (5,0.8550) (6,0.8550) (7,0.8550) (8,0.8550) (9,0.8550) (10,0.8550) (11,0.8550) (12,0.8550) (13,0.8550) (14,0.8550) (15,0.8550) (16,0.8900) (17,0.6400)};
\addplot[color=ecneg, opacity=0.43, line width=0.5pt, mark=none, forget plot] coordinates {(1,0.7600) (2,0.7600) (3,0.7600) (4,0.7600) (5,0.7600) (6,0.7600) (7,0.7600) (8,0.7600) (9,0.7500) (10,0.4100) (11,0.6950) (12,0.7250) (13,0.7300) (14,0.6450) (15,0.6500) (16,0.6500) (17,0.6900)};
\addplot[color=ecpos, opacity=0.48, line width=0.5pt, mark=none, forget plot] coordinates {(1,0.3375) (2,0.3775) (3,0.3950) (4,0.3725) (5,0.3915) (6,0.3955) (7,0.4115) (8,0.4180) (9,0.4185) (10,0.4165) (11,0.3740) (12,0.3675) (13,0.3240) (14,0.2545) (15,0.2640) (16,0.2745) (17,0.9455)};
\addplot[color=ecpos, opacity=0.52, line width=0.5pt, mark=none, forget plot] coordinates {(1,0.0700) (2,0.0700) (3,0.0700) (4,0.0700) (5,0.0700) (6,0.0700) (7,0.0700) (8,0.0700) (9,0.0750) (10,0.0600) (11,0.0550) (12,0.0550) (13,0.0550) (14,0.0550) (15,0.1900) (16,0.1800) (17,0.0300)};
\addplot[color=ecneg, opacity=0.79, line width=0.5pt, mark=none, forget plot] coordinates {(1,0.5550) (2,0.5550) (3,0.5750) (4,0.5750) (5,0.5650) (6,0.5650) (7,0.5550) (8,0.5550) (9,0.5550) (10,0.5550) (11,0.5550) (12,0.5550) (13,0.5550) (14,0.5650) (15,0.3750) (16,0.3850) (17,0.3250)};
\addplot[color=black, line width=1.2pt, mark=*, mark size=1.0pt, forget plot] coordinates {(1,0.3500) (2,0.3950) (3,0.3950) (4,0.3800) (5,0.3850) (6,0.3850) (7,0.3800) (8,0.3800) (9,0.3750) (10,0.3750) (11,0.4300) (12,0.4900) (13,0.5400) (14,0.5450) (15,0.6500) (16,0.7300) (17,0.7900)};
\end{axis}
\end{tikzpicture}

{\footnotesize\textbf{EC-Fusion} top-8:\par}
\vspace{-2pt}
\begin{itemize}[leftmargin=1em, topsep=0pt, itemsep=-2pt, parsep=0pt]
\item {\tiny \textcolor{ecneg}{$r{=}-0.71$} [Politics/Election] Will a Democrat win Nevada US Senate Election?}
\item {\tiny \textcolor{ecpos}{$r{=}+0.72$} [Politics/Election] Will a Democrat win Nevada US Senate Election?}
\item {\tiny \textcolor{ecneg}{$r{=}-0.72$} [Politics/Election] Will a Republican win Nevada US Senate Electi...}
\item {\tiny \textcolor{ecneg}{$r{=}-0.51$} [Politics/Election] Trump gets more black voters than in 2020?}
\item {\tiny \textcolor{ecneg}{$r{=}-0.24$} [Politics/Election] Will a Democrat win Pennsylvania US Senate El...}
\item {\tiny \textcolor{ecpos}{$r{=}+0.33$} [Politics/Election] Will Donald Trump win the popular vote in the...}
\item {\tiny \textcolor{ecpos}{$r{=}+0.39$} [Politics/Election] [Single Market] Will Gavin Newsom win the U.S...}
\item {\tiny \textcolor{ecneg}{$r{=}-0.90$} [Politics/Election] Biden wins the Popular Vote?}
\end{itemize}
\end{minipage}
  \caption{\textbf{Why graph-based retrieval beats text-based retrieval.}
  Same query (``Will Tempo launch a token by December 31 2026?'',
  category: Crypto), two retrieval methods. \emph{Left:} Semantic retrieval
  finds near-duplicate questions in the same category (mean $|r|{=}0.486$
  with the query) --- a textual echo chamber. \emph{Right:} EC-Fusion
  retrieves topically unrelated events (mean $|r|{=}0.567$, all in
  Politics/Election) whose prices nonetheless co-fluctuate with the query,
  including strongly anti-correlated trajectories that are equally
  informative for forecasting. Black: query trajectory; blue/red:
  positive/negative correlation; opacity scales with $|r|$.}
  \label{fig:retrieval_overlay}
\end{figure*}

This appendix expands the qualitative example referenced in Section~\ref{sec:discussion}. The query event is ``\textit{Will Tempo launch a token by December 31, 2026?}'' (Polymarket Crypto, $T_{\text{in}}{=}10$).

\xhdr{What Semantic retrieves}
Semantic retrieval returns eight near-duplicate questions about other crypto token launches: six near-rephrasings of ``Will Berachain launch a token in December?'', ``Will Hyperliquid launch a token in December?'', and the query itself --- all in the Crypto category. The mean absolute Pearson correlation between the query's price trajectory and these retrieved events is only $0.486$, indicating that surface-level textual similarity does not translate into useful predictive signal.

\xhdr{What EC-Fusion retrieves}
EC-Fusion retrieves \emph{zero} crypto events. All eight retrievals are U.S.\ Politics/Election questions (e.g., ``Will a Democrat win Nevada US Senate Election?'', ``Trump gets more black voters than in 2020?'', ``Biden wins the Popular Vote?''), drawn from late-2024 election markets that share a common volatility regime with the query. The mean $|r|$ rises to $0.567$, and several retrievals are strongly anti-correlated ($r{=}-0.71$, $r{=}-0.90$) --- equally informative for the downstream forecasting model.

\xhdr{Component complementarity within EC-Fusion}
The Granger and EC components of EC-Fusion themselves discover almost disjoint candidate sets: on Polymarket, the mean per-query Jaccard overlap between the two components' retrievals is $0.003$, supporting the design choice of an adaptive fusion rather than a redundant ensemble.

\end{document}